\documentclass[useAMS,usegraphicx,usenatbib,referee]{mn2e}
\usepackage{times}

\newcommand{\Rsun}{\ensuremath{\,{\rm R}_\odot}}                        
\newcommand{\kms}{\,km\,s$^{-1}$}                                       

\title[MY\,Serpentis: a high-mass triple system in the Ser OB2 association]
      {MY\,Serpentis: a high-mass triple system in the Ser OB2 association\thanks{Based on the data obtained at T\"{U}B\.{I}TAK National Observatory}}

\author[ \"{O}. \c{C}ak{\i}rl{\i} et al.]
  { C.~Ibanoglu$^1$, \"{O}. \c{C}ak{\i}rl{\i}$^1$\thanks{e-mail:omur.cakirli@gmail.com}, E.~Sipahi$^1$,\\ \\
  $^1$Ege University, Science Faculty, Astronomy and Space Sciences Dept., 35100 Bornova, \.{I}zmir, Turkey\\
  }

\begin{document} \maketitle 
\begin{abstract}
We present spectroscopic observations of the massive multiple system HD\,167971, located in the open cluster NGC\,6604. The
brighter component of the triple system is the overcontact eclipsing binary MY\,Ser with an orbital period of 3.32\,days. The
radial velocities and the previously published UBV data obtained by \citet{may10} and the UBVRI light curves by \citet{dav88}
are analysed for the physical properties of the components. We determine the following absolute parameters: for the primary star 
M$_p$=32.23$\pm$0.54 M$_{\odot}$, R$_p$=14.23$\pm$0.75 R$_{\odot}$; and for the secondary star M$_s$=30.59$\pm$0.53 M$_{\odot}$, 
R$_s$=13.89$\pm$0.75 R$_{\odot}$. Photoelectric times of minimum light are analyzed under the consideration of the light-time
orbit. The center-of-mass of the eclipsing binary is orbiting around the common center-of-gravity of the triple system with 
a period of 21.2$\pm$0.7\,yr and with a projected semi-major axis of 5.5$\pm$0.7\,AU. The mass function for the third star 
was calculated as 0.370$\pm$0.036 M$_{\odot}$. The light contributions of the third star to the triple system in the UBV pass-bands
were derived and the intrinsic magnitudes and colors were calculated individually for the three stars. The components of the eclipsing
pair were classified as O7.5 {\sc iii} and O9.5 {\sc iii}. The intrinsic color indices for the third star yield a spectral type of 
(O9.5-B0) {\sc iii-i}. {\bf This classification leads to constrain the inclination of the third-body orbit, which should be about 30$^{o}$, and 
therefore its mass should be about 29 M$_{\odot}$. MY\,Ser is one of the rare massive O-type triple system at a distance of 
1.65$\pm$0.13\,kpc, the same as for the NGC\,6604 embedded in the Ser\,OB2 association.  }     
\end{abstract}

\section{Introduction}\label{sec:intro}
Massive stars are known as progenitors of supernovae. The end-product of the supernova explosions are neutron stars or 
black holes which can be sources of strong gravitational waves. Therefore massive stars affect not only dynamical but 
also chemical evolution of the galaxies. They ionize the surrounding material and emit UV radiation. Almost at the 
beginning of their life they start to lose considerable amount of their material. Studies of physical characteristics 
of massive components of early-type double-lined eclipsing binaries are essential for understanding of the final 
stages of stellar evolution. The two components of an eclipsing binary system have the same age and chemical 
composition when they born but slightly different masses and radii and temperatures. Therefore, theoretical stellar 
models must be able to match their precisely determined physical properties for one age and chemical composition. This 
comparison for the detached eclipsing binaries leads to improvement in input physics and success of the treatment 
in stellar models of several physical phenomena, for example mass loss, mixing-length, convective core overshooting 
and opacity \citep{cas07}. In addition to this, some hints about the spin and orbital angular momentum exchange in close binaries 
due to mass-transfer and loss will be obtained.  

\citet{zin07} propose that the incidence of binary and multiple stars among the massive O-stars is much larger than 
that for solar-like stars. This difference in multiplicity properties is attributed to the differences in the star 
formation process between massive and low-mass stars. While low-mass stars may loose angular momentum by magnetic- and 
disk-related processes, these are ineffective in massive star formation due to short time scale of formation. \citet{mas98} 
estimate that O-star binary frequency is larger than 60\%, even this may reach to 100 \% as the gap in orbital periods 
between close and wide binaries is filled. They also note that most binaries occur in clusters and associations and that 
binaries are less common among field star and especially among young runaway stars. So far, many eclipsing binary systems 
have been discovered in the open clusters and associations. They not only offer determining the properties of the cluster 
but also allow one to calculate an accurate distance, age and chemical composition as a whole. Accurate physical parameters 
of stars, especially masses, radii and effective temperatures may only be determined from the analyses of multi-passband 
light curves and radial velocities. These analyses may also yield signs about the existence of additional components. 
Eclipsing binaries which are members of physically bounded multiple stellar systems provide additional constraints for 
the reliability of evolutionary models. The physical parameters that can be evaluated for the members of a multiple 
system should be represented by the same isochrone that fit the eclipsing pair. Recently, \citet{tor10} collected 
astrophysical parameters of 94 detached non-interacting eclipsing binaries. Masses and radii of these stars have 
been determined with an accuracy of 3\% or more. This sample includes only three high mass binaries with a component 
as massive as 27 M$_{\odot}$.

The massive binary system MY\,Ser (HD\,167971, V=7.65, U-B=-0.34, B-V=0.75\,mag) is the central star of NGC\,6604, which is 
listed by \citet{hum78} among the visually most luminous O stars in the Galaxy. NGC\,6604 is a young, compact open cluster 
lying at the core of the H {\sc ii} region S54 \citep{geo73}, in the Ser\,OB2 association \citep{for00}. First photoelectric 
photometry made by \citet{hil56} who classified it as O8f in the MK system. Later on \citet{con71} refined this classification 
as O7.5If, and \citet{wal72} as O8Ib(f)p. \citet{yam77} re-classified the star as O8Ibf. Its light variability was detected 
by \citet{lei84}, and later, its eclipsing nature with an orbital period of  3.32 days was revealed by \citet{gen85}. Photometric 
and spectroscopic observations of the star were obtained by \citet{lei87}. They derived the orbital period for the eclipsing 
pair as 3.32 days and classified the stars as O8I+O5-8V. They have also pointed out that a third component of spectral type 
O5-8V dominates the optical and the UV luminosity of the eclipsing pair. First attempt for the solution of the light curve 
is made by \citet{dav88}. They obtained UBVRI light curves and performed preliminary analyses of these curves assuming the 
light contribution of the third star as 63.3\%, 31.55\% and zero. Later on UBV passbands and their analyses are published 
by \citet{may10}. They find light contribution of the third star in the U, B and V light curves as 55.2\%, 56.5\% and 56.7\%, 
respectively. The radial velocities of both components of the eclipsing pair were revealed by \citet{may11} using the ESO 
archive spectra. Combining the results of  the radial velocity and light curves analyses they estimate the masses of the 
components as 37.4 and 34.8 M$_{\odot}$.  
          
In this paper we present new spectroscopic observations and radial velocities of both components of the eclipsing pair. By 
analysing the previously published light curves and the new radial velocities we obtain orbital parameters for the stars. 
Combining the results of these analyses we obtain absolute physical parameters of both components. In addition we reveal 
for the first time orbital parameters of the third-body's orbit and the parameters of the third star.         

\section{Data acquisition}                              
\subsection{Spectroscopy}
Optical spectroscopic observations of the MY\,Ser were obtained with the Turkish Faint Object Spectrograph Camera 
(TFOSC)\footnote{http://tug.tug.tubitak.gov.tr/rtt150\_tfosc.php} attached to the {\bf 1.5\,m telescope in July, 21--29, 2012}, 
under good seeing conditions. Further details on the telescope and the spectrograph can be found at 
http://www.tug.tubitak.gov.tr. The wavelength coverage of each spectrum was 4000-9000 \AA~in 12 orders, with 
a resolving power of $\lambda$/$\Delta \lambda$ $\sim$7\,000 at 6563 \AA~and an average signal-to-noise 
ratio (S/N) was $\sim$120. We also obtained high S/N spectra of four early type standard stars 21\,Cyg, $\tau$\,Her, 
HR\,153 and 21\,Peg for use as templates in derivation of the radial velocities. 

The electronic bias was removed from each image and we used the 'crreject' option for cosmic ray removal. Thus, the 
resulting spectra were largely cleaned from the cosmic rays. The echelle spectra were extracted and wavelength 
calibrated by using Fe-Ar lamp source with help of the IRAF {\sc echelle} package, see \citet{Simkin}.

\subsection{Light curves}
The UBV and intermediate-passband photometric observations have been obtained in 1985 by \citet{lei87} in three 
observatories. Despite some gaps in the resulting light curve the overall-shape of the light curve is well-represented by 
their observations. Depth of the primary minimum is deeper only 0.03 mag and the phase interval between the two eclipses 
is about 0.5, giving an evidence of circular orbit. Almost the same epoch \citet{dav88} performed photometric observations 
in the Cousins-Bessel UBVRI filters. They have published five-passband light curves as well as individual differential 
photometric measurements. The most outstanding feature in their light curves is the asymmetry, in particular, light level 
immediately following primary eclipse appears to be depressed. UBV photometry of MY\,Ser in seven seasons of the years from 
1990 to 1994 is performed by \citet{may10}. Fortunately, one can reach to their original data. In consequence, two 
original multi-passband data sets are available for the use in interpretations.

\section{Analysis of the $O-C$ curve}
\citet{lei87} found from the line spectrum that MY\,Ser is a triple system with two close eclipsing O stars and a distant 
companion also of spectral type O. They note that the third star is the most luminous component of the system both in 
the optical and the UV spectra. The third star has been angularly resolved for the first time by \citet{bec12} using 
multi-epoch {\sc VLTI} observations. Their observations provide direct evidence for a gravitational link between the eclipsing 
pair and the O8 supergiant. They measured separations between the components A and B (the eclipsing binary) vary from 8 
to 15 mas over the three-year baseline. They suggest that the stars evolve on a wide and eccentric orbit which is 
not coplanar with the orbit of the inner eclipsing pair.

Since the photometric observations of the eclipsing pair go back to middle of 1980s the orbit of the eclipsing pair 
around the center-of-mass of triple system can be determined from the times of mid-eclipses. However, the orbital 
period of the eclipsing binary is too long to be obtained a complete minimum in a night in an observing site. The 
eclipse, from first to fourth contact, lasts about 20 hours. For this reason a few times of mid-eclipse was seen 
in the literature. They could be obtained from the observations at the near to mid-eclipse. Therefore we try to 
estimate mid-primary and mid-secondary eclipses using the observations obtained in the ingress and egress of the 
minima. For this reason, we performed a preliminary analysis of the light curves obtained by \citet{may10} and 
obtained the best fit representation of the observed light curves. Plotting the observations with the same scales 
that of the computed curve and determined the minimum time by shifting it along the time axis. We note that the 
shape of the light curve is assumed to be constant during the time base of the observations. As the best fit is 
obtained the time for mid-primary or mid-secondary eclipses are read off from the observations. Using this procedure 
we obtained 20 times for the mid-primary and 21 for the secondary eclipse. The times are given in Table 1 with those taken 
from O-C GATEWAY data-base in the literature. The uncertainties of the minima derived in this study are about a 
few hundredths of a day.

\begin{table}
\caption{Times of mid-eclipses for MY\,Ser and the O-C residuals (see text)}
\label{O-C values}
\begin{center}
\begin{tabular}{crccr}
\hline
HJD-2400000	&	E	&	O-C(I)	&	O-C(II)	&	Ref	\\
 \hline
45555.0000	&	0.0 	&	0.0013	&	0.0123	&	1	\\
45937.0000	&	115.0	&	0.0157	&	0.0264	&	1	\\
46230.9410	&	203.5	&	-0.0063	&	0.0044	&	1	\\
46232.6125	&	204.0	&	0.0044	&	0.0151	&	1	\\
46235.94	&	205.0	&	0.0103	&	0.0210	&	2	\\
46237.58	&	205.5	&	-0.0105	&	0.0002	&	2	\\
48059.47	&	754.0	&	-0.0262	&	-0.0163	&	2	\\
48062.80	&	755.0	&	-0.0178	&	-0.0079	&	2	\\
48067.7894	&	756.5	&	-0.0108	&	-0.0010	&	1	\\
48069.43	&	757.0	&	-0.0310	&	-0.0212	&	2	\\
48071.13	&	757.5	&	0.0082	&	0.0180	&	2	\\
48449.7664	&	871.5	&	-0.0195	&	-0.0098	&	1	\\
48453.11	&	872.5	&	0.0025	&	0.0122	&	2	\\
48501.2700	&	887.0	&	-0.0009	&	0.0087	&	1	\\
48753.7130	&	963.0	&	-0.0007	&	0.0089	&	1	\\
49155.63	&	1084.0	&	0.0009	&	0.0103	&	2	\\
49157.32	&	1084.5	&	0.0301	&	0.0395	&	2	\\
49158.95	&	1085.0	&	-0.0007	&	0.0087	&	2	\\
49163.97	&	1086.5	&	0.0369	&	0.0463	&	2	\\
49459.58	&	1175.5	&	0.0232	&	0.0324	&	2	\\
52083.69	&	1965.5	&	0.0575	&	0.0664	&	2	\\
52088.6490	&	1967.0	&	0.0341	&	0.0422	&	1	\\
52455.68	&	2077.5	&	0.0267	&	0.0374	&	2	\\
52460.66	&	2079.0	&	0.0242	&	0.0359	&	2	\\
52495.54	&	2089.5	&	0.0273	&	0.0359	&	2	\\
52500.52	&	2091.0	&	0.0249	&	0.0359	&	2	\\
52744.69	&	2164.5	&	0.0562	&	0.0589	&	2	\\
52782.88	&	2176.0	&	0.0476	&	0.0589	&	2	\\
52872.54	&	2203.0	&	0.0240	&	0.0343	&	2	\\
52877.53	&	2204.5	&	0.0316	&	0.0344	&	2	\\
52935.63	&	2222.0	&	0.0033	&	0.0112	&	2	\\
53056.87	&	2258.5	&	0.0044	&	0.0112	&	2	\\
53189.76	&	2298.5	&	0.0298	&	0.0357	&	2	\\
53413.97	&	2366.0	&	0.0308	&	0.0371	&	2	\\
53556.76	&	2409.0	&	-0.0086	&	-0.0034	&	2	\\
53571.71	&	2413.5	&	-0.0059	&	-0.0035	&	2	\\
53576.68	&	2415.0	&	-0.0173	&	-0.0098	&	1	\\
53795.93	&	2481.0	&	0.0051	&	0.0080	&	2	\\
53830.80	&	2491.5	&	-0.0018	&	0.0080	&	2	\\
54192.87	&	2600.5	&	0.0121	&	0.0166	&	2	\\
54247.61	&	2617.0	&	-0.0545	&	-0.0441	&	2	\\
54350.62	&	2648.0	&	-0.0146	&	-0.0109	&	2	\\
54355.59	&	2649.5	&	-0.0270	&	-0.0153	&	2	\\
54559.90	&	2711.0	&	0.0037	&	0.0136	&	2	\\
54574.85	&	2715.5	&	0.0064	&	0.0135	&	2	\\
54637.95	&	2734.5	&	-0.0042	&	0.0072	&	2	\\
54672.87	&	2745.0	&	0.0388	&	0.0409	&	2	\\
54888.69	&	2810.0	&	-0.0461	&	-0.0386	&	2	\\
54906.96	&	2815.5	&	-0.0450	&	-0.0386	&	2	\\
54941.87	&	2826.0	&	-0.0120	&	-0.0097	&	2	\\
55129.54	&	2882.5	&	-0.0132	&	-0.0068	&	2	\\
\hline
\end{tabular}
\end{center}
\begin{list}{}{}
\item[Ref:]{\small (1) O-C Gateway, (2) This study}
\end{list}
\end{table}

The O-C(I) residuals given in the third column of Table 1 are computed with the linear light elements,
\begin{equation}
HJD~(Min~I) = 2\ 445\ 554.9987 + 3^{\rm d}.321615 \times E.
\end{equation}
and are displayed graphically against the years in Fig.1. Despite large scatter in the O-C(I) residuals they indicate a 
slow decrease and a steeper increase, then a decrease again. At a first glance the variations of the residuals look like 
a distorted sine curve. Since the residuals for both minima follow almost the same behavior such a variation resembles a 
light-time effect, orbiting around a third-body. The time delay or advance of the observed eclipses caused by the 
influence of a third star can be represented by

\begin{equation}
\Delta T=\frac {a_{12} \sin i_{3}} {c}  \left\{ \frac{1-e_3^2}{1+e_3 \cos \nu_3} \sin[\nu_3 + \omega_3]+e_3 \sin \omega_3 \right\}
\end{equation}
where $a_{12}$ is the semi-major axis of the eclipsing pair's orbit around the barycenter and $i_3$, $e_3$, $\omega_3$ are 
the inclination, eccentricity and longitude of the periastron of the third-body orbit, respectively. The semi-amplitude of the 
light-time effect is A$_{3}$ = $a_{12}$ $\sin i_3$/c, where c is speed of light. The only parameter varying with time 
in Eq.2 is the true anomaly, $\nu_3$, of the third-body. The resulting ephemeris is given by

\begin{equation}
T_{ec}=T_1+EP_1+\Delta T
\end{equation}                                     
{\bf where T$_1$ is the starting epoch, E is the integer eclipse cycle number and $P_1$ is the orbital period of the eclipsing 
binary. First we estimated the initial parameters using the trial-and-error method. Then, a linear least squares fit is applied in order}
to obtain five independent variables A$_3$,T$_3$, P$_3$, $e_3$, 
$\omega_3$ for the third-body orbit plus two variables $T_1$ and $P_1$ for the eclipsing binary. The parameters are given 
Table\,2 with their standard deviations. Subtracting the contribution of the light-time we obtained the O-C(II) residuals 
and plotted in the bottom panel of Fig.\,1. This solution indicates that MY\,Ser revolves around the common center-of-gravity 
with a third-body. The orbit is highly eccentric, amounting 0.53. Using the projected semi-major of 5.5\,AU and a period of 
21.2\,yr the mass function of the third star is calculated as 0.370${\pm}$0.036 M$_{\odot}$.        

{\bf The first O-C analysis confirms that there is a gravitational connection between the MY\,Ser and the third-star as suggested by \citet{bec12}. The radio
light curve obtained by \citet{blo07} reaches to maximum close to 1988-1989 and minimum between 1995 and 1999.} The maximum 
radio flux density to the orbital phase close to periastron passage, when the separation between the eclipsing pair and 
the third-star, is thus confirmed by the O-C analysis of the eclipsing pair.  

\begin{figure}
\center
\includegraphics[width=9cm,angle=0]{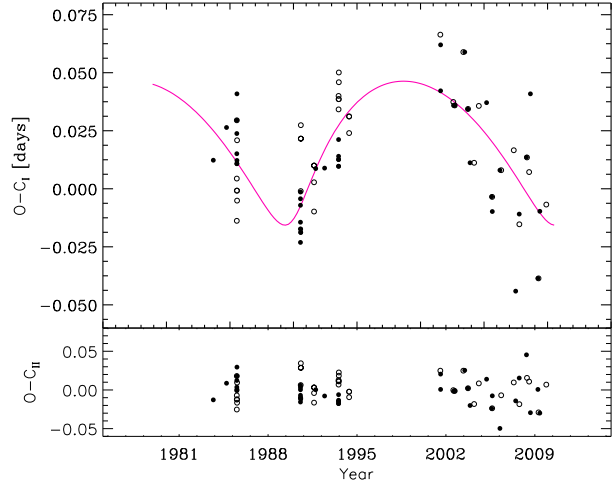}
\caption{Residuals for the times
of minimum light of MY\,Ser with respect
to the linear light elements. The continuous
curve represents the light time effect. In the bottom panel the 
O-C(II) residuals, after subtracting the light-time, are shown. } \end{figure}

\begin{table}
\caption{Orbital parameters for the third component of MY\,Ser .}
\label{table7}
\begin{tabular}{ll}
\hline
Parameter & Value \\
\hline
$T_{1}$ (HJD) & 2445554.9626${\pm}$ 0.0034 \\
$P_{1}$ (d)     & 3.321616${\pm}$ 0.000002 \\
$A_{3}$ (d)      	& 0.032${\pm}$0.005 \\
$e_{3}$      	& 0.53${\pm}$0.05 \\
$w_{3}$			 & 294${\pm}$ 30 \\
$a_{12}sini$ (AU)	& 5.5${\pm}$ 0.7 \\
$P_{3}$ (yr)& 21.2${\pm}$ 0.7 \\
$T_{3}$ (HJD) & 2455592${\pm}$ 547 \\
$\Sigma (O-C)^{2}$ & 0.0167 \\
 \hline
\end{tabular}
\end{table}

\section{Radial Velocities}
{\bf To derive the radial velocities, the eight spectra obtained for the system are cross-correlated 
against the template spectra of standard stars on an order-by-order basis using the {\sc fxcor} 
package in IRAF \citep{Simkin}. 

The spectra showed three cross-correlation peaks in the quadratures, one for each component 
of the triple system. Thus, the peaks are fitted independently with a $Gaussian$ profile to measure the 
velocities and their errors for the individual components. If the three peaks appear blended, a triple 
Gaussian was applied to the combined profile using {\it de-blend} function in the task. For each of 
the eight observations} we then determined a weighted-average radial velocity for each star from all 
orders without significant contamination by telluric absorption features. Here we used as weights 
the inverse of the variance of the radial velocity measurements in each order, as reported by {\sc fxcor}.

{\bf The heliocentric radial velocities for the primary (V$_p$), the secondary (V$_s$) components and also for the third star (V$_t$) are 
listed in Table\,3,} along with the dates of observations and the corresponding orbital phases computed 
with the new ephemeris given above. The velocities in this table have been corrected to the heliocentric 
reference system by adopting a radial velocity value for the template stars.{\bf The radial velocities are 
plotted against the orbital phase in Fig.\,2, together with the error bars.} We analysed all the 
radial velocities for the initial orbital parameters using the {\sc RVSIM} software program \citep{kane}. 
Figure\,2 shows the best-fit orbital solution to the radial velocity data. The results of the analysis 
are as follow: $\gamma$= -14.5$\pm$3.2 \kms, $K_1$=265$\pm$4 and $K_2$=279$\pm$4 \kms with circular 
orbit. Using these values we estimate the projected orbital semi-major axis and mass ratio as: 
$a$sin$i$=35.72$\pm$0.54 \Rsun~ and $q=\frac{M_2}{M_1}$=0.949$\pm$0.028.

{\bf The radial velocities of the third star vary from -4 to -43 \kms. Since the observations were obtained in successive 
eight nights, these changes should be arisen from the measurement errors, due to the blending of the
three lines. The average radial velocity is determined as -26$\pm$13 \kms.}

\begin{figure}
\includegraphics[width=8.5cm,angle=0]{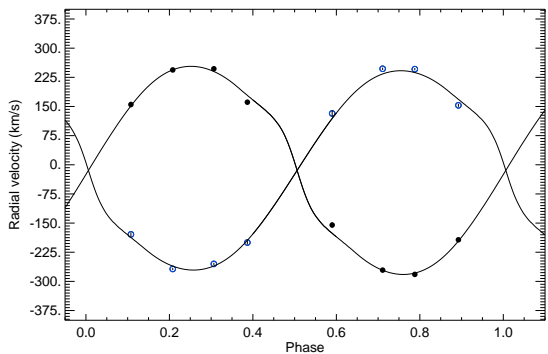}
\caption{Radial velocities for the components of MY\,Ser. Symbols 
with error bars show the radial velocity measurements for the 
components of the system (primary: open circles, 
secondary: filled circles). } \end{figure}

\begin{table}
\scriptsize
\centering
\begin{minipage}{85mm}
\caption{{\bf Heliocentric radial velocities of MY\,Ser and the third star. The columns give the heliocentric 
Julian date, the orbital phase (according to the ephemeris in \S 3), the radial velocities of 
the three components with the corresponding standard deviations.}}

\begin{tabular}{@{}ccccccccc@{}cc}
\hline
HJD 2400000+ & Phase & \multicolumn{2}{c}{Star 1 }& \multicolumn{2}{c}{Star 2 } & \multicolumn{2}{c}{Star 3 }	\\
             &       & $V_p$                      & $\sigma$                    & $V_s$   	& $\sigma$	&V$_t$& $\sigma$\\
\hline
56130.3678  &   0.8034  &    227    &   5   &   -275&   3   &-4&10\\
56131.3096	&	0.0869	&	-155	&	7	&	132	&	7	&-37&5\\
56132.3140	&	0.3893	&	-173	&	3	&	153	&	4	&-41&9\\
56133.3625	&	0.7050	&	244	    &	4	&	-278&	3	&-27&8\\
56135.2888	&	0.2849	&	-272	&	4	&	246	&	3	&-43&7\\
56136.3513	&	0.6048	&	155	    &	7	&	-199&	5	&-7&11\\
56137.2779	&	0.8837	&	161	    &	6	&	-200&	5	&-18&8\\
56138.3544	&	0.2078	&	-271    &	5	&	247	&	3	&-30&6\\
\hline
\end{tabular}
\end{minipage}
\end{table}

\begin{figure}
\center
\includegraphics[width=8.5cm,angle=0]{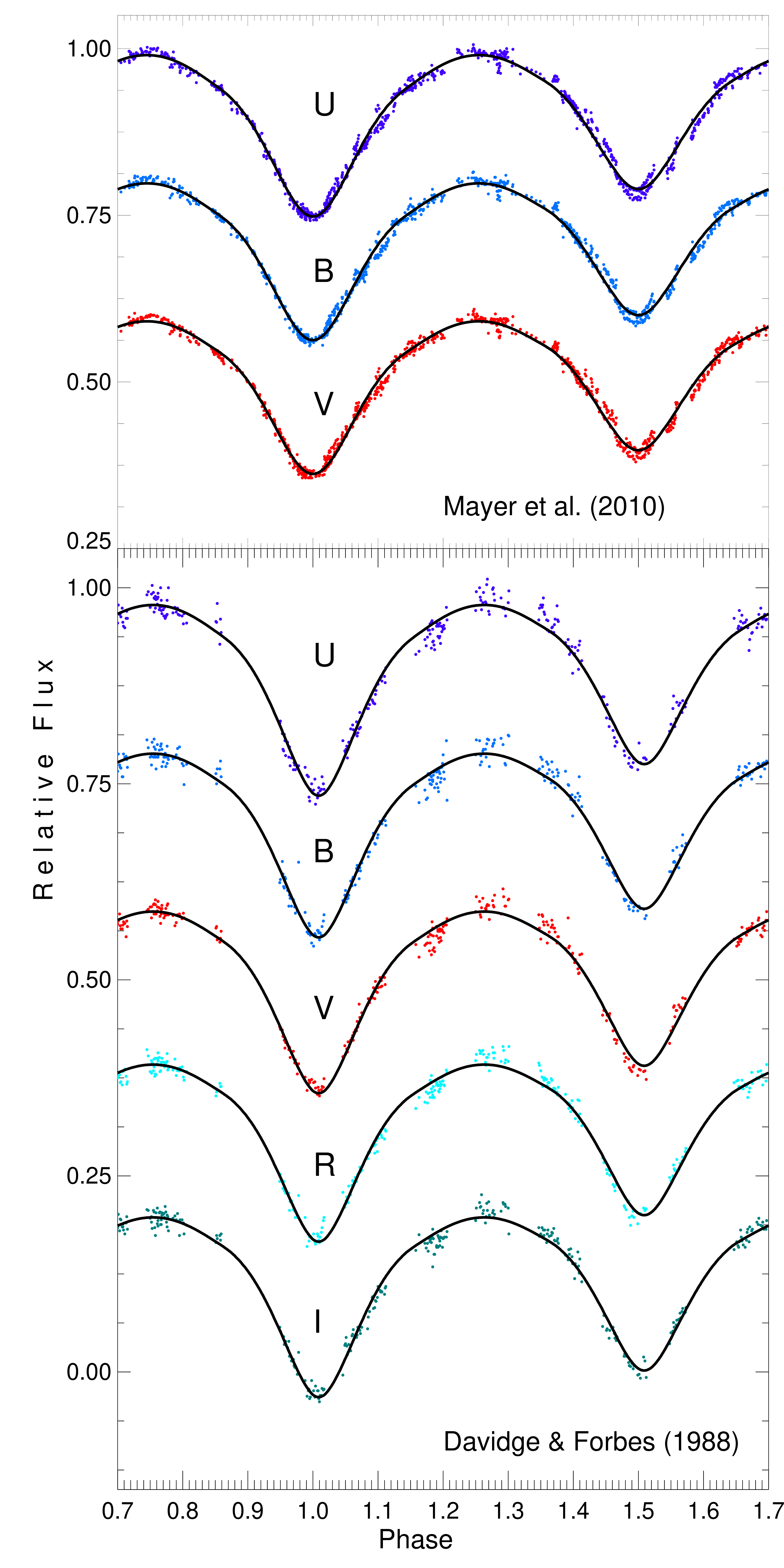}
\caption{a) The Mayer et al.'s UBV, b) Davidge and Forbes' UBVRI light curves of MY\,Ser. The continuous lines show the best-fit model.} \end{figure}

\section{Light curves and their analyses}
{\bf As we quoted in \S1 the optical photometric studies of} MY\,Ser were made by \citet{lei87} in the UBV, by 
\citet{dav88} in the UBVRI and by \citet{may10} in the UBV. The original data of the latter two studies were published 
for use of other investigators. \citet{dav88} obtained 1\,520 points in the five passbands. The standard deviations of 
differential brightnesses were given as 0.014\,mag in U, 0.005\,mag in B, and 0.006\,mag in V, R, and I. Although there 
are some gaps the shape of the light curve was revealed. The UBV photometric observations of \citet{may10} were obtained 
between 1990-1994, which  contain 3\,042 points. Their three passbands light curve are almost complete. They did not give 
standard deviations of their observations. We estimate accuracies of differential observations from the maxima about 
0.010 in U, 0.009 in B,and 0.010\,mag in V-passband. The light curves are shown in Fig.\,3.         

Most commonly used code for modeling the light curves of the eclipsing binaries is that of \citet{Wilson71}. 
This code was up-dated and implemented in the {\sc phoebe} code of \citet{prs05}. One of the main difficulties
in this modeling is determination of individual effective temperatures of both stars. Generally used practice is 
to estimate effective temperature of primary star and determine that of the secondary star. Effective temperature of 
the primary star could be estimated from its spectra or color indices. However, spectrum of an eclipsing binary is a 
composite spectra of both stars. In some cases decomposition of the spectra is almost impossible. Similarly the color
indices are also include colors of both stars. If the stars of an eclipsing binary have similar properties, i.e. their
temperatures and luminosities are very close, it is difficult task to estimate their contribution to the total 
luminosity of the system. 

The average colors are measured by \citet{lei87} as U-B=-0.34$\pm$0.02, B-V=0.75$\pm$0.01, V-R=0.52$\pm$0.01, V-I=1.05$\pm$0.01\,mag. 
They point out that the colors remain constant independently of the phase. In addition, infrared colors J-H=0.205$\pm$0.05,
H-K=0.177$\pm$0.05\,mag are given in the 2MASS catalog \citep{cut03}. The quantity $Q$=($U-B$)-($E_{(U-B)}$/($E_{(B-V)}$)$(B-V)$ 
is independent of interstellar extinction. The average value obtained is ($E_{(U-B)}$/($E_{(B-V)}$)=0.72$\pm$0.03 (\citep{joh53};\citep{hov04}). 
We compute the reddening-free index as $Q$=-0.88$\pm$0.03. The values of index were calculated by \citet{hov04} begining 
from O8 to G2 spectral type for the luminosity classes between main-sequence and supergiants. As pointed out by \citet{joh53}
this quantity is almost constant between O8 and B0. Both calibrations yield an approximate spectral type of the triple system as O9.
A preliminary analysis of the light curves obtained by \citet{may10} yields log g=3.6 for both stars of the eclipsing pair which 
indicates that they should be luminosity class of III, i.e. giants. Using the calibration of \citet{mar05} for the stellar parameters 
of galactic O stars we obtained a spectral type of O(7.5$\pm$0.5)III for the more massive and O(9.5$\pm$0.5) III for the less massive secondary
component of MY\,Ser. The effective temperatures are estimated as 34\,000$\pm$1\,000K and 30\,000$\pm$1\,000K for the primary and 
secondary stars, respectively. Thus, the intrinsic color of $(B-V)_0$=-0.30$\pm$0.01 mag and the interstellar reddening 
$E_{(B-V)}$=1.05$\pm$0.01\,mag are estimated.       

Logarithmic limb-darkening coefficients were interpolated from the tables of \citet{ham93}. They are updated at every 
iteration. The gravity-brightening coefficients $g_1$=$g_2$=1.0 and albedos $A_1$=$A_2$=1.0 were fixed for both 
components, as appropriate for stars with radiative atmospheres. Since the preliminary analysis indicates that both 
stars fill their corresponding Roche lobes synchronous rotations were adopted and Mode 6 was used in the solution. 
{\bf This mode is used for the contact systems, e.g., both stars fill their limiting Roche lobes.} The 
adjustable parameters in the light curves fitting were the orbital inclination $i$, the effective temperature of the 
secondary star T$_{eff_2}$, the luminosity of the primary L$_1$, the third-body light contribution $l_3$ and the 
zero-epoch offset. The parameters of our final solution are listed in Table\,4. The uncertainties assigned to the 
adjusted parameters are the internal errors provided directly by the code. In the last three lines of Table\,4 sums 
of squares of residuals $\sum$$(O-C)^{2}$, number of data points $N$, and standard deviations $\sigma$ of the observed 
light curves are presented, respectively. The computed light curves are compared with the observations in Fig.\,3. We 
also applied Mode\,3, suitable for the over contact binaries, the orbital parameters did not vary considerably but the 
sum of squares of the residuals is slightly increased.  

\citet{wil92} suggestion about the unit of third light was to express it in the light curve of the triple system
at a definite orbital phase. The values of $l_3$ in units of total triple system light were estimated at the 
reference phase 0.25, from the Mayer et al.'s light curves, to be 0.388$\pm$0.002, 0.395$\pm$0.002 and 0.406$\pm$0.002 
in U, B, V passbands, respectively. The light contributions of the third star from the  Davidge and Forbes' light 
curves are found to be  0.472$\pm$0.006, 0.483$\pm$0.005, 0.485$\pm$0.006, 0.495$\pm$0.005, and 0.484$\pm$0.005 
in U, B, V, R and I passbands. Its contribution increases towards the longer wavelengths except the I-passband. Our spectra 
show that the light contribution of the third star to the total light is about 0.39 at the $\lambda$ 6563. It is
in a good agreement with that obtained from the analysis of Mayer et al.'s light curves.

\begin{table}
\scriptsize
\caption{Results of the simultaneous analyses of the Mayer's $UBV$, and Davidge \& Forbes's $UBVRI$ light curves for MY\,Ser.}
\begin{tabular}{lrr}
\hline
Parameter & Mayer et al. (UBV)&Davidge \& Forbes (UBVRI)  \\
\hline	
$i^{o}$			               			 					&73.76$\pm$0.08		&77.98$\pm$.30							\\
T$_{eff_1}$ (K)												&34\,000[Fix]		&34\,000[Fix]									\\
T$_{eff_2}$ (K)												&29\,350$\pm$50		&29\,750$\pm$115							\\
$\Omega_1$=$\Omega_2$										&3.667$\pm$0.021	&3.667$\pm$0.021						\\
$r_1$														&0.3825$\pm$0.0016	&0.3825$\pm$0.0016					\\
$r_2$														&0.3733$\pm$0.0017	&0.3733$\pm$0.0017				\\
$\frac{L_{1}}{(L_{1}+L_{2})}$U  							&0.578$\pm$0.002	&0.570$\pm$0.006					\\
$\frac{L_{1}}{(L_{1}+L_{2})}$B 							&0.569$\pm$0.002	&0.563$\pm$0.006					\\
$\frac{L_{1}}{(L_{1}+L_{2})}$V							&0.568$\pm$0.002	&0.562$\pm$0.006					\\
$\frac{L_{1}}{(L_{1}+L_{2})}$R 							&					&0.562$\pm$0.007					\\
$\frac{L_{1}}{(L_{1}+L_{2})}$I							&					&0.561$\pm$0.006					\\
$\frac{L_{3}}{(L_{1}+L_{2}+L_{3})}$U 			 			&0.388$\pm$0.002	&0.472$\pm$0.006					\\
$\frac{L_{3}}{(L_{1}+L_{2}+L_{3})}$B  						&0.395$\pm$0.002	&0.483$\pm$0.005					\\
$\frac{L_{3}}{(L_{1}+L_{2}+L_{3})}$V  						&0.406$\pm$0.002	&0.485$\pm$0.006					\\
$\frac{L_{3}}{(L_{1}+L_{2}+L_{3})}$R 						&                 	&0.495$\pm$0.005					\\
$\frac{L_{3}}{(L_{1}+L_{2}+L_{3})}$I 						&	                  &0.484$\pm$0.005					\\
$\sum(O-C)^{2}$													&0.2176				&0.2127               \\	
$N$															&3\,042				&1\,520											\\	
$\sigma$													&0.0085				&0.0118				\\				
\hline
\end{tabular}
\end{table}

\section{Results and discussion}
Combining the results of radial velocities and light curves analyses we have calculated the absolute parameters 
of the stars. Separation between the components of eclipsing pair is calculated as  $a$=36.69$\pm$0.20R$_{\odot}$. The 
fundamental stellar parameters for the components such as masses, radii, luminosities are listed in Table\,5 
together with their formal standard deviations. The standard deviations of the parameters have been determined by 
JKTABSDIM\footnote{This can be obtained from http://http://www.astro.keele.ac.uk/$\sim$jkt/codes.html} code, which 
calculates distance and other physical parameters using several different sources of bolometric corrections 
\citep{sou05}. The mass for the primary of $M_p$ = 32.23 $\pm$ 0.54M $_{\odot}$ and secondary of $M_s$=30.59$\pm$0.53M$_{\odot}$ 
are consistent with O(7-8)III and O9.5III stars.

\begin{table}
\scriptsize
 \setlength{\tabcolsep}{2.5pt} 
  \caption{Properties of the MY\,Ser components}
  \label{parameters}
  \begin{tabular}{lrr}
  \hline
   Parameter 																& Primary	&	Secondary										\\   
   \hline
    Mass (M$_{\odot}$) 															& 32.23$\pm$0.54 	& 30.59$\pm$0.53		\\
   Radius (R$_{\odot}$) 														& 14.23$\pm$0.75	& 13.89$\pm$0.75			\\   
   $T_{eff}$ (K)																		& 34\,000$\pm$1000	& 29\,350$\pm$1000		\\
   $\log~(L/L_{\odot})$															& 5.383$\pm$0.069		& 5.102$\pm$0.054		\\
   $\log~g$ ($cgs$) 																& 3.640$\pm$0.046 	& 3.638$\pm$0.047			\\     
   $(vsin~i)_{calc.}$ (km s$^{-1}$)										& 214$\pm$11				& 210$\pm$11					\\    
\hline  
  \end{tabular}
\end{table}

We estimate the interstellar reddening of $E$(B-V)=1.05\,mag which is consistent with that of 1.02\,mag obtained 
by \citet{bar00} for the cluster NGC\,6604. Using this value and the apparent visual magnitude of the triple system 
as V=7.65 given by \citet{lei87} and taking into account their contributions we find apparent visual magnitudes 
of the three stars as $V_p$=8.83,	$V_s$=9.13 and $V_3$=8.63\,mag. The bolometric corrections are adopted from 
\citet{mar05} as -3.20, -2.80 for the primary and secondary stars, respectively. In Table\,6 we present magnitudes 
corrected for interstellar absorption, color indices of the three stars as well as the derived distances for 
the components of the eclipsing binary system. Absolute bolometric magnitude of the Sun is taken as 4.74. The 
distances for the primary and secondary star found to be 1\,645$\pm$133 and 1\,640$\pm$143 pc, respectively. The 
distance to the NGC\,6604 has been subjected to the various studies. \citet{mof75} derived a distance of 1.64\,kpc, 
while \citet{bar00} estimated a distance of 1.7\,kpc. Our result is consistent with the earlier values. The distance 
determination is, of course, depended upon the total apparent magnitude of the multiple system. It varies from 7.37 
(\citet{may10}) to 7.65\,mag. \citet{lei87}. If we adopt the former the distance to the system is reduced to 1\,445\,pc, 
decreased by about 12\%. 

Fortunately, the third star in the HD\,167971 system has been angularly resolved by \citet{bec12}. Their multi-epoch 
VLTI observations provided evidence for a gravitational link between the third star and the close eclipsing pair. The 
separation between the third star and the eclipsing binary has been varied from 8 to 15\,mas over the three-year 
baseline. Although limited interferometric measurements are available they suggest that the components evolve 
on a wide and very eccentric orbit, probably $e$ $\ge$ 0.5. Using near infrared luminosity ratio of 0.8 they 
suggest spectral types for the stars in the close binary between O6 and O7 main-sequence stars. As a results they 
assumed an (O6.5V+0.6.5V)+O8I spectral classification for the stars in the HD\,167971. 

In \S3 we presented parameters of the third-body orbit. Since the O-C curve was however only poorly covered by minimum 
times and the times were very uncertain the parameters given in Table\,2 were not well established. Even a preliminary 
set of parameters for the third-body orbit is obtained the parameters appear to support the interferometric 
measurements. The orbital period about the center-of-gravity with the third-star is derived as 21.2$\pm$0.7 yr, 
$a_{12}$ $\sin i_3$=5.5$\pm$0.7 AU and $e$=0.53$\pm$0.05. Using the projected semi-major axis and the period of 
the third-body orbit we estimate a mass function for the third star as $f(m)$=0.370$\pm$0.036M$_{\odot}$. Since 
the masses of the eclipsing pair's components are derived as 32.23 and 30.59 solar masses and the eclipsing pair orbiting around 
common center-of-mass with a period of 21.2\,yr the mass of the third star can easily be calculated from the mass 
function. The mass of the third body has been computed as 12.70, 15.0 and 28.8 M$_{\odot}$ for the inclination of 
90, 60 and 30 degrees, respectively. A plausible inclination of the third-body should be smaller than 30 degree when 
the light contribution is taken into account. Assuming an inclination of 30 degrees we estimate separation between 
the third-body and eclipsing pair as 8.07 and 26.26\,AU, when they are at the periastron and apastron on their orbits, 
respectively. They correspond to linear angular separations of 4.9 and 15.9\,mass for a distance of 1\,650 pc. They
are in good agreement with the interferometric measurements between 8.7 and 15.1\,mas, obtained by \citet{bec12}. 
Unfortunately, the four interferometric measurements were made in two epochs, e.g. in 2008 and 2011. We derived the 
semi-amplitude of the radial velocity of the eclipsing pair's gravity center as K$_{12}$=9$\pm$1 \kms. Therefore, the 
semi-amplitude of the radial velocity of the third star should be about 20\,\kms. {\bf This result is in a good agreement
with the measured average radial velocities of -26$\pm$13 \kms for the third star.}

The roughly inferred intrinsic magnitudes and color indices are already presented in Table\,6. With the color indices 
of (U-B)$_0$=-1.080 and (B-V)$_0$=-0.275\,mag the third star appears to be later spectral type than both components of 
the eclipsing binary. Its intrinsic colors yield a spectral type close to B0 giant or supergiant. If it is a (O9.5-B0)III star 
its absolute visual magnitude and bolometric correction are deduced as -5.10 and -2.84\,mag from the calibrations of 
\citet{mar05}. Using the intrinsic apparent visual magnitude of 5.36 we obtained a distance to the third star as 
1\,280 pc. If it were a supergiant, i.e. (O9.5-B0)I, its distance would be 2\,130\,pc. For an extreme case, if the 
third star is assumed to be a main-sequence star then it would lie at a distance about 7\,00\,pc. These comparisons 
point out that the third star's luminosity class is between giant and supergiant. It could also be explained by a 
(O9.5-B0)III-I star at the same distance as the binary. 

{\bf The three lines are well separated at the quadratures of the eclipsing pair. We measured equivalent widths (EW) of the third star's
He {\sc i } 4471 and He {\sc ii} 4541 lines. The average logarithmic ratio of the equivalent widths (EW) of the third star's is about 1.19$\pm$0.17, which corresponds to 
a B0-type star \citep{huc96}. On the other hand the ratio between the Si {\sc iv} 4088 and He {\sc i} 4143 lines 
is obtained as 0.34$\pm$0.04. This value corresponds to III-I luminosity class stars. The ratios of the
selected spectral lines confirm our spectral classification from the intrinsic color indices.
}

\citet{bec12} estimated spectral types for the stars in close binary O6.5V, assuming that luminosity ratios should 
be similar for both bolometric and K-band luminosities. They also noted that for a giant luminosity class the close 
binary components should not be earlier than O8 spectral type. They propose for an O8I for the third-star. On the 
other hand, \citet{lei87} reported that the O supergiant is the most luminous and dominates the optical and UV 
spectra. However, our analysis of the \citet{may10} light curves show that luminosity ratios are 0.63, 0.65 and 0.68 
in the U, B and V-passband. These ratios are obtained as 0.89, 0.94, 0.94, 0.98 and 0.94 from the UBVRI light curves of 
\citet{dav88}. In contrary to the \citet{bec12}, we find that eclipsing binary is brighter than the third 
star. The same result is also confirmed by the spectra we obtained around the H$_{\alpha}$. Perhaps, future spectroscopy will reveal more about its possible nature.  

NGC\,6604 is located in a rich part of the Milky Way in the constellation $Serpens$. Its galactocentric distance is 
about 6.9\,kpc, which places it on the outer boundary of the $Carina-Sagittarius$ arm, about 65\,pc above the galactic 
plane. \citet{rei08} suggest that NGC\,6604 forms the densest part of the wider Ser\,OB2 association, which contains 
about 100 OB stars. His studies show that the age of the cluster is about 4-5\,Myr and star formation is still ongoing 
in the association.

\begin{figure*}
\center
\includegraphics[width=15cm,angle=0]{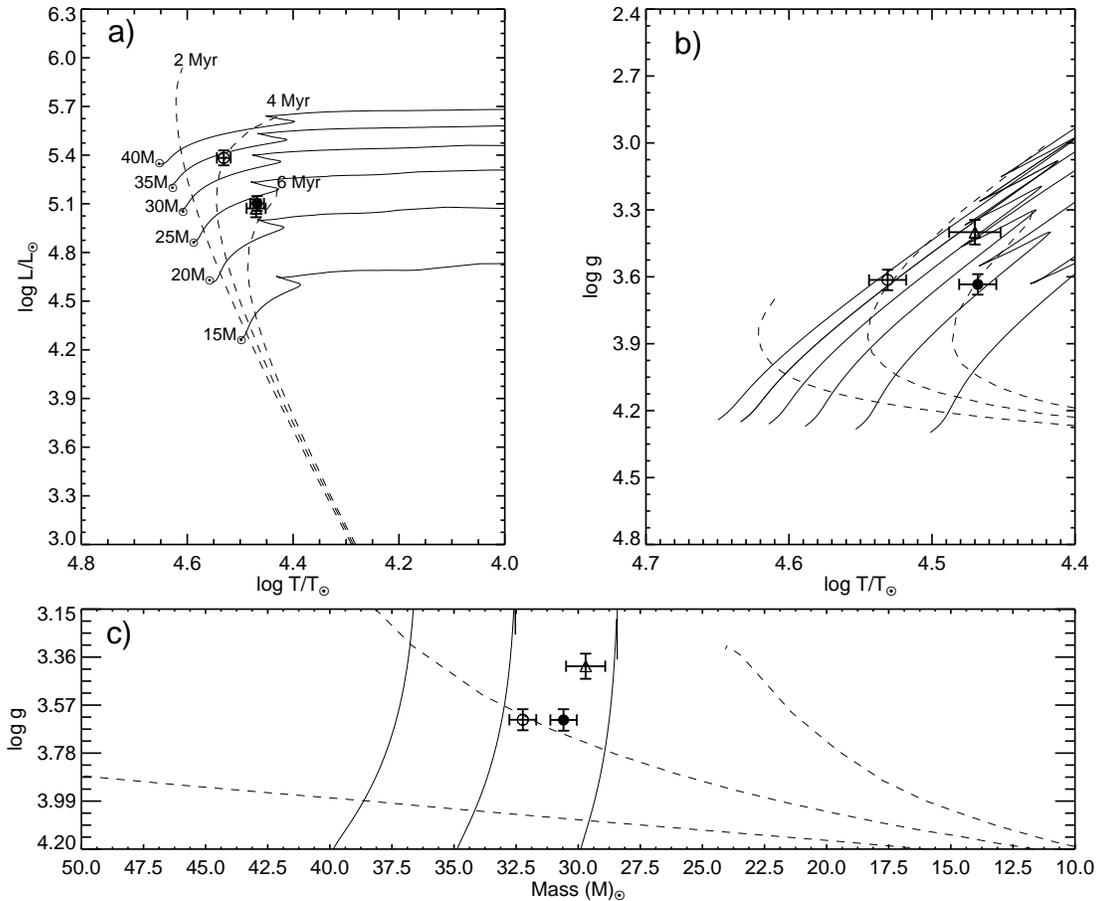}
\caption{Locations of the components of MY\,Ser on the 
a) log T$_{eff}$-log L/L$_{\odot}$, 
b) log T$_{eff}$-log g and 
c) mass-log g  planes. Evolutionary tracks and isochrones are adopted from \citet{eks12}
for $Z$=0.014 with mass-loss, labelled in age (Myr). {\bf Location of the third star on the log T$_{eff}$-log L/L$_{\odot}$ diagram is also shown at the panel (a,b, and c) with
triangle symbol.}
} \end{figure*}

Figure\,4 shows the components of MY\,Ser in the log T$_{eff}$-log L/L$_{\odot}$, log T$_{eff}$-log g and mass-log g 
planes. The evolutionary tracks and isochrones for the $Z$=0.014 with mass-loss are adopted from \citet{eks12}. While 
the location of the more massive companion in the T$_{eff}$-log L/L$_{\odot}$ and log T$_{eff}$-log g is consistent with 
the models for single stars with an age of about 4\,Myr the secondary star appears to lower temperature with respect 
to its mass with an age of about 6\,Myr. {\bf Three components of the massive multiple system HD\,167971 seem to have 
same age about 4\,Myr.} These age 
estimates are consistent with that proposed by \citet{rei08} for the open cluster NGC\,6604. \citet{hil96} discussed 
the properties of six O to B0 detached binaries. They concluded that the masses and radii for the components in these 
systems are all in agreement with the theoretical models of solar composition. However, O-star temperature should be 
reduced by an average of about 1\,000\,K to obtain in agreement with the models. In the case of MY\,Ser the secondary 
star, in contrary to their result, seems to have lower temperature by about 3\,000\,K.

MY\,Ser appears to have just filling the entire outer contact surface with fill-out factor of 0.99. As suggested by 
\citet{may10}, who determined a fill-out factor of 0.50, MY\,Ser is an overcontact binary.

\begin{table}
\scriptsize
 \setlength{\tabcolsep}{2.5pt} 
  \caption{ The intrinsic apparent visual magnitudes, color indices, spectral and distances for the components of MY\,Ser. The intrinsic magnitudes and color 
  indices for the third-star and HD\,167971 stellar system are given in coloumns 4 and 5, respectively.}
  \label{parameters}
  \begin{tabular}{lrrrr}
  \hline
   Parameter 																& Primary	&	Secondary	&Third-body		&Total      	\\   
   \hline
    U ($mag$) 															& 4.109$\pm$0.00 	& 4.449$\pm$0.00	&4.007$\pm$0.00		&2.980$\pm$0.00	\\
    B ($mag$) 															& 5.237$\pm$0.00 	& 5.539$\pm$0.00	&5.087$\pm$0.00		&4.079$\pm$0.00	\\ 
    V ($mag$) 															& 5.563$\pm$0.00 	& 5.859$\pm$0.00	&5.362$\pm$0.00		&4.383$\pm$0.00	\\ 
    U-B ($mag$) 														& -1.128$\pm$0.00 	& -1.090$\pm$0.00	&-1.080$\pm$0.00	&-1.099$\pm$0.00	\\ 
    B-V ($mag$) 														& -0.326$\pm$0.00 	& -0.320$\pm$0.00	&-0.275$\pm$0.00	&-0.304$\pm$0.00	\\
    Sp.Type 														& O(7.5$\pm$0.5) III 	& O(9.5$\pm$0.5) III	&---	&---\\
	M$_{bol}$  															& -8.718$\pm$0.00 	& -8.015$\pm$0.00	&	---	&		---\\  	
	BC ($mag$) 															& -3.20$\pm$0.00 	& -2.80$\pm$0.00	&	---	&	---	\\  
    M$_V$ ($mag$) 														& -5.518$\pm$0.00 	& -5.215$\pm$0.00	&---		&		---\\     
    V-M$_V$ ($mag$)														& 11.081$\pm$0.00 	& 11.074$\pm$0.00	&	---	&	---	\\    
  	$d$ (pc)															& 1645$\pm$133	 	& 1640$\pm$143		&---		&	---	\\
\hline  
  \end{tabular}
\end{table}

\section{Conclusion}
HD\,167971 is one of the rare triple system with massive components. Analysis of the radial velocities and light
curves suggest that the inner binary consists of an O7.5 III and O9.5 III stars. The spectral type of the third star 
is estimated as O9.5 between giant and supergiant luminosity classes. The times for mid-primary and mid-secondary 
eclipses are analysed for the first time for the parameters of the third-body orbit. The gravitational link between the 
MY\,Ser and a third-star is confirmed.

Using our estimation of $E$(B-V) and A$_v$ we determined a distance of about 1\,650\,pc which is in agreement with the 
distance of the open cluster NGC\,6604. The properties of the components of MY\,Ser are compared with theoretical 
models. While the locations of the more massive primary component on the effective temperature-luminosity, effective 
temperature-gravity and mass-gravity planes are consistent with the models the less massive secondary star appears to 
have lower temperature with respect to its mass. Both components of MY\,Ser locate on the mass-gravity plane, that 
matched very well with an isochrone of 4\,Myr. The ages suggested for NGC\,6604 range from 4 to 6\,Myr. Our results 
yield that the age of the close binary, is close to 4\,Myr, and thus NGC\,6604, where star formation still is going on.

\section*{Acknowledgments}
We thank to T\"{U}B{\.I}TAK National Observatory (TUG) for a partial support in using RTT150 
telescope with project number 11BRTT150-198.
We also thank to the staff of the Bak{\i}rl{\i}tepe observing station for their warm hospitality. This study 
is supported by Turkish Scientific and Technology Council under project number 112T263.
The following internet-based resources were used in research for this paper: the NASA Astrophysics Data System; the 
SIMBAD database operated at CDS, Strasbourg, France; T\"{U}B\.{I}TAK ULAKB{\.I}M S\"{u}reli Yay{\i}nlar 
Katalo\v{g}u-TURKEY; and the ar$\chi$iv scientific paper preprint service operated by Cornell University. 
{\bf The authors are indebted to the anonymous referee for his/her valuable suggestions which improved the paper.}

\bibliographystyle{elsarticle-harv}

\begin{thebibliography}{00}

%
\bibitem[Aparicio et al.(2007)]{cas07} Aparicio, A., Hidalgo, S.~L., Gallart, C., \& Cassisi, S.\ 2007, IAU Symposium, 241, 267 

\bibitem[\protect\citeauthoryear{Barbon et al.}{2000}]{bar00} Barbon R., Carraro G., Munari U., Zwitter T., Tomasella L., 2000, A\&AS, 144, 451 

\bibitem[\protect\citeauthoryear{Blomme et al.}{2007}]{blo07} Blomme R., De Becker M., Runacres M.~C., van Loo S., Setia Gunawan D.~Y.~A., 2007, A\&A, 464, 701 

\bibitem[\protect\citeauthoryear{Conti \& Alschuler}{1971}]{con71} Conti P.~S., Alschuler W.~R., 1971, ApJ, 170, 325 

\bibitem[Cutri et al.(2003)]{cut03} Cutri, R.~M., Skrutskie, M.~F., van Dyk, S., et al.\ 2003, VizieR Online Data Catalog, 2246, 0 


\bibitem[\protect\citeauthoryear{Davidge \& Forbes}{1988}]{dav88} Davidge T.~J., Forbes D., 1988, MNRAS, 235, 797 

\bibitem[\protect\citeauthoryear{De Becker et al.}{2012}]{bec12} De Becker M., Sana H., Absil O., Le Bouquin J.-B., Blomme R., 2012, MNRAS, 423, 2711 

\bibitem[Ekstr{\"o}m et al.(2012)]{eks12} Ekstr{\"o}m, S., Georgy, C., Eggenberger, P., et al.\ 2012, A\&A, 537, A146 

\bibitem[Forbes(2000)]{for00} Forbes, D.\ 2000, AJ, 120, 2594 

\bibitem[Georgelin et al.(1973)]{geo73} Georgelin, Y.~M., Georgelin, Y.~P., \& Roux, S.\ 1973, A\&A, 25, 337 

\bibitem[Hilditch et al.(1996)]{hil96} Hilditch, R.~W., Harries, T.~J., \& Bell, S.~A.\ 1996, A\&A, 314, 165 

\bibitem[\protect\citeauthoryear{Hiltner}{1956}]{hil56} Hiltner W.~A., 1956, ApJS, 2, 389 

\bibitem[\protect\citeauthoryear{Hovhannessian}{2004}]{hov04}Hovhannessian R.~K., 2004, Ap, 47, 499 

\bibitem[\protect\citeauthoryear{Humphreys}{1978}]{hum78} Humphreys R. M., 1978, ApJ, 38, 309 .

\bibitem[\protect\citeauthoryear{Johnson \& Morgan}{1953}]{joh53} Johnson H.~L., Morgan W.~W., 1953, ApJ, 117, 313 

\bibitem[\protect\citeauthoryear{Kane, Schneider, \& Ge}{2007}]{kane} Kane S.~R., Schneider D.~P., Ge J., 2007, MNRAS, 377, 1610



\bibitem[\protect\citeauthoryear{Leitherer et al.}{1984}]{lei84} Leitherer C., Stahl O., Zickgraf F.~J., Klare G., Wolf B., 1984, IBVS, 2539, 1 

\bibitem[Leitherer et al.(1987)]{lei87} Leitherer, C., Forbes, D., Gilmore, A.~C., et al.\ 1987, A\&A, 185, 121 
 

\bibitem[\protect\citeauthoryear{Martins, Schaerer,\& Hillier}{2005}]{mar05} Martins F., Schaerer D., Hillier D.~J., 2005, A\&A, 436, 1049 

\bibitem[\protect\citeauthoryear{Mason et al.}{1998}]{mas98} Mason B.~D., Gies D.~R., Hartkopf W.~I., Bagnuolo W.~G., Jr., ten Brummelaar T., McAlister H.~A., 1998, AJ, 115, 821 

\bibitem[\protect\citeauthoryear{Mayer et al.}{2010}]{may10} Mayer P., Bo{\v z}i{\'c} H., Lorenz R., Drechsel H., 2010, AN, 331, 274 

\bibitem[\protect\citeauthoryear{Mayer, Drechsel \& Bro{\v z}}{2011}]{may11} Mayer P., Bro{\v z} H., Lorenz R., Drechsel H., 2011, From Interacting Binaries to Exoplanets: Essential Modeling Tools, Proc. IAU Symp. 282, 311

\bibitem[Moffat \& Vogt(1975)]{mof75} Moffat, A.~F.~J., \& Vogt, N.\ 1975, A\&AS, 20, 155 

\bibitem[Pr\v{s}a \& Zwitter(2005)]{prs05} Pr\v{s}a A., Zwitter T. 2005, ApJ, 628, 426

\bibitem[Reipurth(2008)]{rei08} Reipurth, B.\ 2008, Handbook of Star Forming Regions, Volume II, The Southern Sky, p.590 

\bibitem[\protect\citeauthoryear{Simkin}{1974}]{Simkin} Simkin, S. J., 1974, A\&A, 31, 129 

\bibitem[Southworth et al.(2005)]{sou05} Southworth J., Smalley B., Maxted P. F. L., Claret A. \& Etzel P. B. 2005, MNRAS, 363, 529

\bibitem[\protect\citeauthoryear{Torres, Andersen \& Gim{\'e}nez}{2010}]{tor10} Torres G., Andersen J., Gim{\'e}nez A., 2010, A\&ARv, 18, 67 

\bibitem[\protect\citeauthoryear{van Genderen}{1985}]{gen85} van Genderen A.~M., 1985, IBVS, 2760, 1 

\bibitem[van Hamme(1993)]{ham93} van Hamme, W. 1993 AJ, 106, 2096 

\bibitem[\protect\citeauthoryear{van der Hucht}{1996}]{huc96} {\bf van der Hucht K.~A., 1996, LIACo, 33, 1 }

\bibitem[\protect\citeauthoryear{Walborn}{1972}]{wal72}Walborn N.~R., 1972, AJ, 77, 312

\bibitem[Wilson \& Devinney (1971)]{Wilson71} Wilson R.E. \& Devinney E.J., 1971, ApJ, 166, 605

\bibitem[\protect\citeauthoryear{Wilson}{1992}]{wil92} Wilson R.~E., 1992, Documentation of Eclipsing Binary Computer Model (Univ. of Florida) 

\bibitem[\protect\citeauthoryear{Yamashita\& Nariai}{1977}]{yam77} Yamashita Y., Nariai K., 1977, "An Atlas of Representative Stellar Spectra", Uni.Tokyo Press

\bibitem[\protect\citeauthoryear{Zinnecker \& Yorke}{2007}]{zin07} Zinnecker H., Yorke H.~W., 2007, ARA\&A, 45, 481 
\end{thebibliography}



\end{document}